\documentclass{appolbnh}
\usepackage{graphicx}
\newcommand{\beq}{\begin{equation}}
\newcommand{\eeq}{\end{equation}}

\begin{document}
\title{{\em{Discovery of High-momentum Nucleon Correlations in  Nuclei --- an Early History}}}
\headtitle{Discovery of High-momentum Nucleon Correlations in  Nuclei \dots}
\author{{L.}{Frankfurt}$^\mathrm{a,b}$,
{M.}{Strikman}$^\mathrm{b,}$\thanks{Corresponding author: \texttt{mxs43@psu.edu}}
\address{$^\mathrm{a}$Tel Aviv University, Tel Aviv, 69978, Israel\\    
$^\mathrm{b}$104 Davey Lab, The Pennsylvania State University\\
University Park, PA 16803, USA\\}}
\headauthor{L. Frankfurt, M. Strikman}
\maketitle
\begin{abstract}
We  summarize the reasons for using the light-cone mechanics of nuclei for description of high-energy nuclear processes 
and describe the steps, which have led to the discovery of high-momentum correlations in nuclei.
\end{abstract}
%\appbDOI

\section{Preface}

%{\it M.S.}
\subsection*{Some personal memories of M.S.}
I  first met Mitya in the fall of 1966 at the Leningrad  University soon after 
we started our undergraduate studies in the Physics Department. We graduated in the spring of 1972 and were admitted to the  ``aspirantura'' in April (my appointment started  two weeks after Mitya due to some bureaucratic clearance games).

Perhaps our most intensive interactions were during working on translation of Richard Feynman's book {\em Photon--Hadron Interactions}.
Vladimir Shechter managed to get a copy of the book a few days earlier than people from the Moscow institution and we got the contract.
While doing the translation, we discussed the Feynman picture a lot.

My interests focused more and more on high-momentum effects in the description of nuclei, while Mitya first worked on topics 
in perturbative high-energy QCD and proclaimed after the first data from the CERN $p\bar p$ collider that nothing of importance could come from this direction of research. I~guess that his opinion about the study of nuclei was even more pessimistic (it was shared practically by all theorists studying strong interactions).
Mitya's attitude started to change in the mid-2000s --- he agreed that the discussion of the nuclear structure at the resolution scale   
of sponta-\break 

\newpage\noindent
neously broken chiral symmetry is important, and he explored the structure of nuclear matter in a~chiral model with an explicit inclusion of ${\mit\Delta}$ isobars in the modeling of nuclear matter.

\vspace{-3mm} 
%{\it L.F.}
\subsection*{Some personal memories of L.F.}
By  the end of the 1960s, it  became evident to me that a deep puzzle exists in particle physics. On the one hand, the 
quark model explained many static properties of hadrons and their interactions. On the other hand,
this hypothesis is inconsistent with what existed at the time of preQCD quantum field theories: due to the Landau nullification of the running coupling constant. The heavy nucleus looked as a good testing ground, especially due to high matter  densities in configurations where nucleons come close to each other. So, I became interested in how these phenomena manifest themselves in high-energy processes,
especially since it became evident that the theoretical method used in nuclear physics contradicted the Feynman parton model. 

Hence, we  decided that it would be beneficial for readers to have  a brief summary of theoretical and experimental discoveries  of nucleons with high momenta and correlations between such nucleons.
We do not discuss spacial correlations between nucleons, which are much more difficult (practically impossible) to observe~\cite{Bethe:1971xm}. However, we use the same name for both of them --- short-range correlations. Our presentation will be mostly chronological --- step by step.

\vspace{-2mm}
\section{Steps  toward  theoretical description and experimental observation  of SRC}

\vspace{-1mm}
\subsection{Step 0. Pre-1974 world}

In 1974, quarks were considered by most as merely a convenient approach to deep inelastic scattering with little connection to hadronic physics (a notable exception were relations between meson--baryon and baryon--baryon total cross sections~\cite{Levin}).   It was suggested for some time that to resolve the microscopic structure  of nuclei, one needs to study scattering at a sufficiently large momentum transfer. Nuclei were considered in nuclear physics as best described by quasiparticle approaches~\cite{Migdal}, in which short-range correlations (SRC) could be 20\% or 50\% --- it does not matter for observables. Interaction between nucleons was due to meson exchanges  for  nuclear model practitioners, and ill-defined starting at distances $1/m_{\pi}$ for groups with high-energy physics background. 
What  helped a bit in our interactions with several people about quark-related topics was the November revolution of 1974 --- the discovery of $J/\psi, {\mit\psi}'$, dynamics of final states in  $e^+e^-$ annihilation into hadrons. (For some time L.F.\ spent more of his time on the charm/charmonium  physics than on physics described here).

\newpage
A good illustration of this common thinking was our experience with submission to \textit{Physics Letters  B} a paper in which we suggested
determining the origin of nuclear forces by means of neutrino--nucleus scattering.  The  referee stated that the  authors suggest yet another way to observe  meson currents and that it would be unfair to encourage experimentalists to spend time on such experiments.  When  we  told James Bjorken  during his visit to Leningrad in the spring of 1975 that the quark exchange was probably more realistic than the meson exchange, he responded --- what is the difference? Our response was that there would be extra antiquarks in nuclei. He was happy and was telling people that this was one of the most interesting things he had learned during the visit to our institute.  It took another 25 years to check experimentally using  the Drell--Yan process that antiquarks are not enhanced in the relevant $x \sim 0.1$--0.2 range~\cite{DY}.

Principle questions we were interested in included  :
\begin{itemize}
\item[\textit{(i)}] Could short-range structure of deuteron and heavier nuclei be more complicated than a collection of nucleons?
It looked like there was no reason to think of a proton and a neutron coming close together as two nucleons since quarks could knead into compact configurations. 
We suggested to name such configurations as  the ``kneaded state'' --- later on, several groups started to refer to such configurations  as   six-quark states.
\item[\textit{(ii)}] Do meson exchanges provide an adequate framework for description of nuclear forces?
\item[\textit{(iii)}] How to test the  hypothesis that nuclear forces are due to meson exchanges?
\item[\textit{(iv)}] What is the dynamics of nuclear forces at internucleon distances $\le  1$~fm, where the wave functions of two nucleons overlap significantly.
\end{itemize}

\subsection{Step 1. The starting point of the time line is 1974}

It was clear that to resolve such configurations in nuclei, one needs to study processes with energy-momentum scales much larger than the typical mass difference between two nucleon and six-quark states. Also, we suggested half jokingly that {\it asymptotic freedom prevents collectivization of  quarks} in line with  our everyday experience.
The effectiveness of this logic was confirmed by a series of experiments at BNL and JLab, 
which directly observed  short-range correlations (SRC) in a series of nuclei, and established similarity of 
the SRC in the deuteron and heavier nuclei with $pn$ correlations giving dominant contribution.

\newpage   
In the fall of 1975, we came across
the paper of West~\cite{West}, which stated that there exists a correction  to the impulse approximation for the scattering off a deuteron.  Such a correction would imply presence in the  deuteron  of hidden (unobservable) configurations\footnote{Several groups reproduced our result~\cite{Westcom} and one or two were  asking for our comments. In response to one of the letters in addition to the discussion of the calculation, M.S.\ cooked up a story that in our department, there is a saying ``Whatever way you calculate a Feynman diagram, if you do it correctly the answer would be the same.''  A member of our department,  whose public service was to serve as a kind of  a censor of correspondence  to be sent abroad  (though we assumed that this was a pure formality), has asked us how it is that he never heard this saying.}.
                           
It became clear that in order  to resolve the short-range structure of nuclei at the level of nucleon/hadronic constituents, one needs processes, which transfer to the nucleon constituents both energy ($q_0$) and momentum ($\vec q\,$) larger than the scale of the 
nucleon--nucleon ($NN$) short-range correlations
\begin{equation}
q_0\, \ge  1\ {\rm GeV} \,, \qquad  ~\vec{q}\;~  \ge 1 \  {\rm GeV} \,.
\end{equation}

\begin{figure}[h!]
\centerline{%
\includegraphics[width=9.5cm]{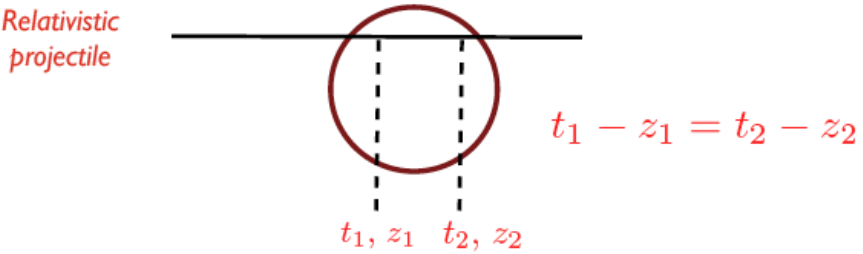}}
\caption{Sketch illustrating LC dominance in high-energy processes.}
\label{fig:f1k}
\end{figure}

Consequently, one needs to describe the processes in the relativistic domain.  The  price to pay is the need to treat the nucleus wave function using light-cone (LC) quantization as the projectile is propagating along the LC: $t-z= \mathrm{const.}$ (see Fig.~\ref{fig:f1k})~\cite{FS81,FS88}. Similarly to perturbative QCD, the cross sections of high-energy processes are expressed through the LC wave functions of nuclei, which satisfy baryon charge, electric charge, and momentum conservation  (all of which were violated in the  calculation of West~\cite{West}).     
Also, the behavior of the cross section near the kinematical  boundary is determined by the wave function at momenta $\to \infty$ in the nucleus wave function.
These requirements cannot be  satisfied in all covariant approaches. 

In Fig.~\ref{fig:f1k}, we sketch the space-time picture of the interaction with a~nuclear target: although the nucleus is at rest, the slice of the wave function along the LC  is selected by the presence of a high-energy projectile.

\vspace{2mm}
%rys.1

\newpage
The LC quantization is uniquely selected in high-energy processes if one tries to express the cross section through elementary amplitudes near the energy shell. 
For example, one can consider the breakup of the deuteron in the impulse approximation:
 $h+D\to X+N$, for $E_h \to \infty$.
It is easy to demonstrate that the difference of the invariant energies in the intermediate $h N$ state  and the final produced system 
$X$ tends to $\infty$ for all quantization axes except the LC one along the beam direction~\cite{FS88}.
Note that in many cases, it is convenient to use two variables to characterize the LC wave function: the  transverse momentum of the constituent and its light-cone fraction. The factor of two in the denominator (factor of $A$ for heavier nuclei) ensures that the wave function is maximal for all nuclei for $\alpha = 1$
\begin{equation}
\alpha = 2 (E_N-p_{3N}) /m_D \,.
\end{equation}

Among other things this  definition of $\alpha$ ensures that the cross section automatically  tends to zero  for $\alpha =2$.

\subsection{Step 2. Deuteron LC wave function}
 
The relation between the LC and non-relativistic (NR) wave functions is quite simple in the case of the two-body system.
The starting point for deriving LC-NR correspondence is a decomposition of the deuteron wave function over hadronic states
\begin{equation}
\left|D\right>=\left|NN\right>+\left|NN\pi\right>+\left|{\mit\Delta} {\mit\Delta}\right>+ \left|NN\pi\pi\right> + \dots \,.
\end{equation}
It would be useless if too many states were involved in the Fock representation. To address the question of the essential degrees of freedom, we cannot use  the experience of the models of $NN$ interactions based on the meson theory of nuclear forces since in such models the  Landau pole close to the mass shell is present and, hence,  a lot of multi-meson configurations are generated.
At the same time,  we can use the information on $NN$ interactions at energies below a few GeV and the chiral dynamics combined with the following general quantum mechanical (QM) principle --- relative magnitudes of different components in the wave function should be similar to that in the $NN$ scattering at the energy corresponding to the off-shellness of the components~\cite{FS81}. Important simplification comes from the composition of the final states in $NN$ interactions: direct pion production is suppressed for a wide range of energies due to the chiral properties of the $NN$ interactions~\cite{FS88}. As a result, the lowest-mass significant components are single and double ${\mit\Delta}$ isobar production of which only ${\mit\Delta}{\mit\Delta}$ is allowed in the deuteron channel. (In the current NR calculations, these components are hidden in the effective two-pion exchange potential.)
As a result, the lowest-energy gap is $2 m_{\mit\Delta} - 2m_N\approx \mbox{600 MeV}$ corresponding to the mass/energy gap for two nucleons with momenta
\begin{equation}
k_N=\sqrt{m_{{\mit\Delta}}^2-m_N^2} \approx 800 \ {\rm MeV} \,.
\end{equation}
For the case of $NN$  interaction in the isospin 1 channel, where a single ${\mit\Delta}$ can be produced, the corresponding momentum 
$k_N\sim 550$ MeV. 

\begin{figure}[htb]
\centerline{%
\includegraphics[width=12.5cm]{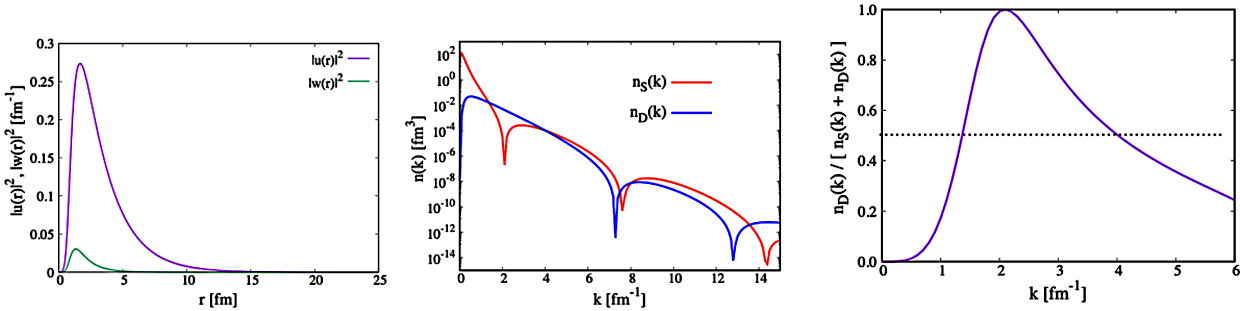}}
\caption{Deuteron $S$- and $D$-wave functions in coordinate and momentum representations.}\vspace{3mm}
\label{fig:fig2}
\end{figure}

Note here that the naive expectation based on experience with smooth potentials that characteristic momenta are inversely proportional to typical distances does not work in the case of a rapidly changing potential like in the deuteron case.
For example, the deuteron $D$-wave is much smaller than the $S$-wave for $r_{12} < 0.5$ fm, while the momentum space wave function is dominated by the $D$-wave for most of the studied momentum range, $300 < k < 800$~MeV, see, \eg, Fig.~\ref{fig:fig2}.
A comparison of the $S$- and $D$-wave  in coordinate space explains also why it is easier to observe momentum  correlations than spacial correlations. In fact, the measurement of the $S/D$ ratio as a function of momentum $k$ makes hard reactions with polarized deuteron the best place to look for the difference between predictions of non-relativistic  and relativistic approaches~\cite{core,FS81}.

%rys.2

\begin{figure}[htb]
\centerline{%
\includegraphics[width=10cm]{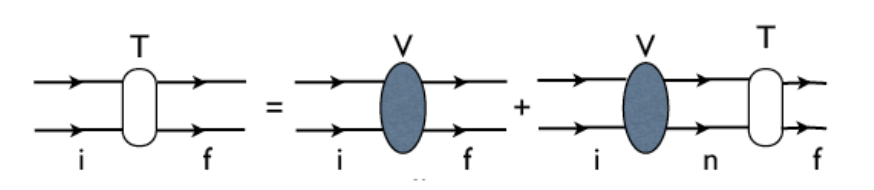}}
\caption{Schematic representation of the LC equation for $NN$ scattering.}
\label{fig:fig3}
\end{figure}

The presence of a small parameter (inelasticity of $NN$ interactions) justifies the use of the relativistic two-nucleon approximation for the LC wave function of the deuteron.  
The first step is to include interactions which do not have two-nucleon intermediate states into kernel $V$ (like in non-relativistic QM) and build a Lippmann--Schwinger-type LC  equation symbolically depicted  in Fig.~\ref{fig:fig3}\footnote{This equation is often referred to as the Weinberg equation. However, in his paper~\cite{Weinberg}, he considered the effects of pions in the wave on function with not fixed number of degrees of freedom approximation.  He also did not discuss the angular condition.}.

%rys.3

\newpage
In this equation, intermediate states contribute to
the LC energy denominators as $1/(p_{n_+} - p_{f_+})$ and the potential $V$ depends both on the light-cone fractions $\alpha_i$ of interacting nucleons and their transverse momenta, $p_\mathrm{t}$ (we also  extended  this approach to the case of a few nucleon systems~\cite{modern}).

The second step is to impose the 
condition that the master equation should lead to the Lorentz invariant on-energy-shell amplitude of $NN$ scattering. 
We have demonstrated~\cite{FS88,Lech}
that to satisfy this requirement, it is {\it necessary} and {\it sufficient} that the
potential $V$ depends on the scalar products of two three-dimensional vectors
$\vec{k}_{i}=(k_{3i},k_{\mathrm{t}i}), \vec{k}_{f}=(k_{3f},k_{\mathrm{t}f})$,
which are related to the LC variables as 
\begin{equation}
\alpha={\sqrt{m^2+k^2} + k_3\over  \sqrt{m^2+k^2} } \,,
\label{alphak}
\end{equation}
where we use normalization of the LC fractions $\sum_{i=1}^{i=A} \alpha_i=A$.
It has to be compared to the relation between the momentum of a proton in the deuteron rest frame
\begin{equation}
{\alpha= \left(\sqrt{m_N^2+p^2 } - p_z\right)\Big/(m_D/2)} \,.
\end{equation}

For small momenta $k$, the $k/p$ ratio is very close to one; for large $k$, the relation between $k$ and $p$ is highly non-linear. 
For example, for 180 degrees, the maximal backward momentum in the deuteron rest frame is $p= 3/4 m_N$, while 
 $k \to \infty$.
 
It is convenient to introduce a single-nucleon LC density of the deuteron $\rho_D^N(\alpha, p_\mathrm{t}) $ normalized using the baryon charge sum rule
 \begin{equation}
 \int \rho^N_D (\alpha, p_\mathrm{t}) {\mathrm{d}\alpha\, \mathrm{d}^2 p_\mathrm{t} \over \alpha} =1 \,.
 \end{equation}
 In the impulse approximation, the process goes as follows. The projectile inelastically interacts, \eg, with a neutron with momentum $p_1$   of the target, releasing a proton with momentum $p_2 = - p_1$ (destroying the potential acting between the neutron and proton), leading to
 \begin{equation}
 {1 \over  \sigma_\mathrm{inel}^{hN} } {\mathrm{d}^{h+D \to N +X }\over \mathrm{d}\alpha /\alpha \,  \mathrm{d}^2 p_\mathrm{t}} = \rho^N_D (\alpha, p_\mathrm{t}) \,,
 \label{deut}
 \end{equation}
 where $\rho(\alpha, p_\mathrm{t})$ is proportional to the non-relativistic nuclear density calculated for $k$ and $\alpha$ according 
 to Eq.~(\ref{alphak}). Similar expression is valid for the case of scattering off few nucleon SRCs~\cite{FS81}.

There are corrections to the impulse approximation but they are on the scale of 30\%.

Other directions of future theoretical research include finding effective probes of the  three-nucleon short-range correlations 
in hard processes and looking for non-nucleonic degrees of freedom in nuclei, in particular, ${\mit\Delta}$ isobars.
Actually, there is very little data on the basic reaction,  where the spectator mechanism dominates and $\alpha >1.3$.

\subsection{Step 3. Discovery of the high-momentum component in nuclei}
 
\begin{figure}[htb]
\centerline{%
\includegraphics[width=8cm]{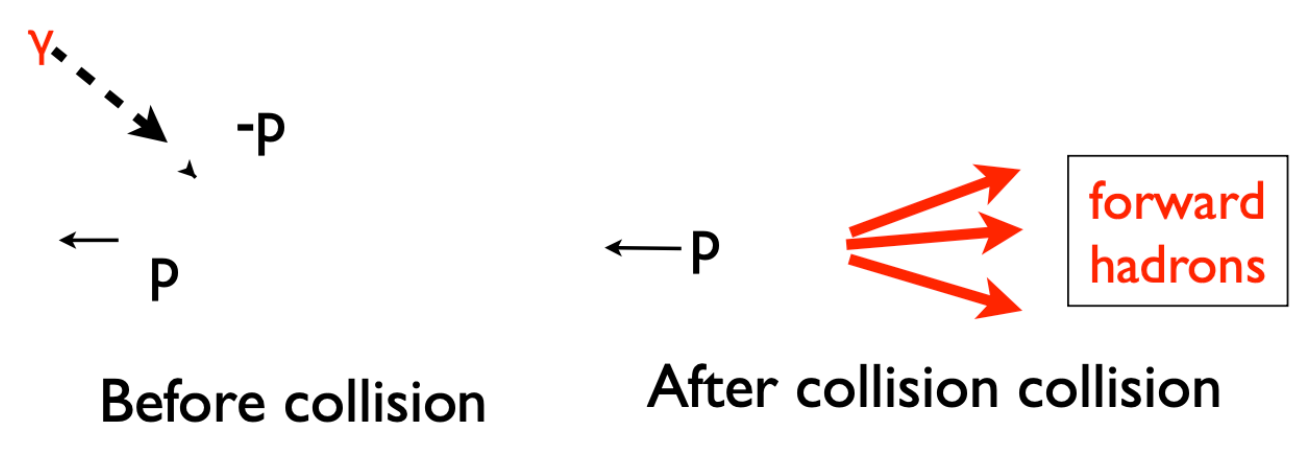}}                            
\caption{Fast backward nucleon production --- the spectator mechanism.}
\label{fig:fig4}\vspace{2mm}
\end{figure}

We were asked back in 1976 at the Rochester Conference in Tbilisi by a~friend of ours --- Valerii Khovansky --- an experimentalist from Moscow who studied $\nu Ne$ scattering using a large FNAL  bubble chamber filled with neon at that time: ``I see protons emitted backward  in our  bubble chamber. Any idea where they  could be coming from?''  We knew the production mechanism for the deuteron target --- hitting a forward nucleon and releasing a nucleon, which was balancing the knockout forward nucleon (including hadrons produced $\nu N$ interaction), see Fig.~\ref{fig:fig4}.

%rys.4

For an $NN$ potential decreasing  as a power of momentum, we expect a similar momentum dependence in different nuclei. So, in analogy with the deuteron breakup, we expected that the proton backward spectra should be proportional 
\begin{equation}
 {1\over A}{\mathrm{d}\sigma^{h+A \to N +X }\over \mathrm{d}\alpha /\alpha \,  \mathrm{d}^2 p_\mathrm{t}} \bigg/
% \over 
  {\mathrm{d}\sigma^{h+D\to N +X }\over \mathrm{d}\alpha /\alpha   \mathrm{d}^2 p_\mathrm{t}}  = a_2(A) \,,
  \label{src2} 
\end{equation}
where $a_2(A)$ denotes the ratio of probabilities for a  nucleon to belong to a~pair correlation in a nucleus and in the deuteron.
We used Eq.~(\ref{deut}) to deter-\break

\newpage\noindent
mine the momentum distribution of protons in nuclei. The only normalized data, which were available in 1976,  were $\gamma $-carbon data at $E_\gamma \sim  3$ GeV 
for a broad range of backward angles~\cite{Kim}\footnote{Data on the backward production  were reported at about the same time by the group of Stavinsky in JINR Dubna, see references~in~\cite{Stavinsky} and the group of Leksin  in ITEP  Moscow (see references in~\cite{Leksin}), but the data were taken at a more limited range of angles.  Also, no tables of cross sections were provided.}.

\begin{figure}[htb]
\centerline{%
\includegraphics[width=7cm]{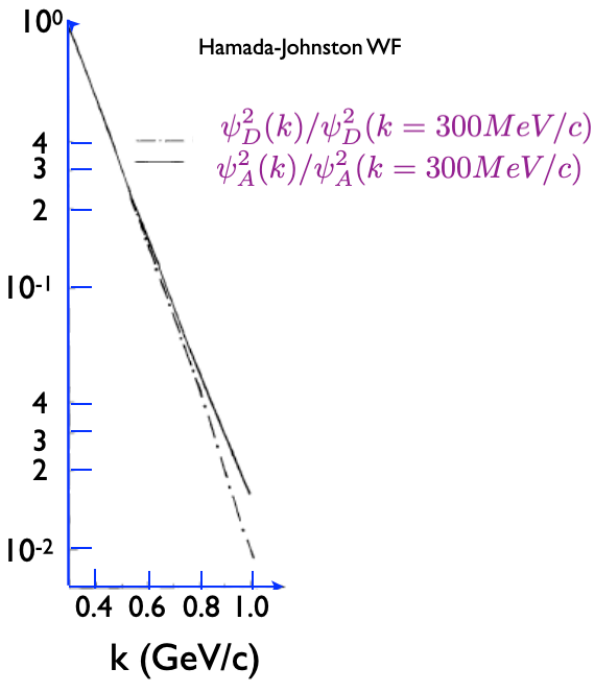}}
\caption{Momentum dependence of the nucleon momentum distribution extracted in~\cite{2Ndisc} from the $\gamma+C \to p+ X$ data~\cite{Kim} for momenta $\geq$  300 MeV.}
\label{fig:fig5} 
\end{figure}

We used Eq.~(\ref{deut}) to extract from the data the momentum distributions in the nucleus at $p \ge 0.3$~GeV/$c$ and found it to be close to the momentum distributions in the deuteron calculated with realistic wave functions,
see Fig.~\ref{fig:fig5}.
Hence, we argued that backward  nucleon production is a good way to study SRC. 

%rys.5

We  submitted the paper to the \textit{Physics Letters B} nuclear section early in 1977 and quickly received a response  from  the nuclear section editor Claude Mahaux. He wrote that his policy is to reject without peer review  all manuscripts claiming  that it is possible to study SRC, since such claims ultimately were turning out wrong.  Luckily, the editor of the particle section of \textit{Physics Letters} --- Peter Landshofff just published a paper reproducing our analysis of the West correction and he quickly accepted the paper.

\newpage
Actually, we would have agreed with Claude Mahaux, if he had added that he applies the  ban to the processes with a few hundred MeV transfer. The early experimental studies were focusing on the photo- and pion absorption. While evidence was
found that absorption by two nucleons dominates, 
no direct link between the process
and scattering off correlated high momentum pairs was established. In fact, the main mechanism
of photoabsorption --- dipole interaction --- involves simultaneous interaction with two nucleons. In
particular, lack of absorption in the $pp$ channel originates not from small probability $pp$ SRC like in the high-energy processes, but from the absence of the dipole interaction. It was also argued that multistep
processes, like the production of ${\mit\Delta}$ isobar in the intermediate state, could not be suppressed.

\subsection{Step 4. Observing SRC in {$(e,e^{\prime})$}{(e,e')} reactions at {$Q^2 \sim \mbox{few GeV}^{\;2}$}{Q2 sim few GeV2}}

The quantities which enter into description of $(e,e^{\prime})$ and $(e,e^{\prime}p)$ reaction cross sections are spectral functions  of nuclei describing the probability to find the $A-1$ system with a given momentum $\vec{p}$ and a large excitation energy~$E$~\cite{FS81}
\begin{equation} \left< E \right >\approx p^2/2m_N \,.
\end{equation}
The model based on this picture~\cite{CiofidegliAtti:1991mm} 
%vg(FS88Ciofi) 
describes well the results of direct calculations of the spectral function.
This excitation energy in the case of SRC kinematics is much larger  than  the one assumed in the $y$-scaling models, which were popular in the seventies and eighties.

\begin{figure}[htb]
\centerline{%
\includegraphics[width=8cm]{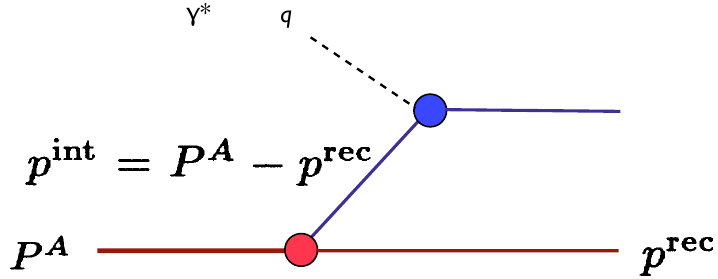}}
\caption{Impulse approximation diagram for the $(e,e^{\prime})$ reaction in the quasi-elastic large-$Q^2$ kinematics.}
\label{fig:fig6}
\end{figure}

The next step in the study of SRC  was to study the $(e,e^{\prime})$ process at  large $x=Q^2/2q_0m_N > 1$ and $Q^2 \ge 1.5$ GeV$^2$. By considering the kinematics of the process depicted in Fig.~\ref{fig:fig6}, one can demonstrate that in the discussed limit, the cross section depends only on the LC fraction $\alpha_\mathrm{int}$ carried by the struck nucleon 
\begin{equation}
\alpha_\mathrm{int}=  (E_\mathrm{int} -\vec{p}_\mathrm{int}\cdot \vec{q}/\left|q\right|) / (M_A/A) \,.
\end{equation}

%rys.6

\newpage\noindent
For intermediate $Q^2$, one has to include the recoil mass effects, which under the assumption of the dominance of two-nucleon correlations allows one to calculate $\alpha_\mathrm{int}= \alpha(x,Q^2)$.
This leads to an expectation~\cite{FS88,Frankfurt:1993sp} of the scaling of the ratios 
of $eA$ and $eD$ cross sections
\begin{equation}
{\sigma_{eA}\left(x,Q^2\right)\over \sigma_{eD}\left(x,Q^2\right)}
_{ \left | 1.3 \le \alpha_\mathrm{int} \le 1.6 \right.}
= {2 \over A}  a_2(A)  \,,
\end{equation}
which  agrees well with the SLAC data~\cite{Frankfurt:1993sp}
 and more recent JLab data. Here, $a_2(A)$ is  the ratio of the light-cone density matrices of the nucleus and the deuteron in the SRC kinematics.
 
Remarkably, $a_2(C)$ we found from the backward nucleon production 
%vg\cite{}  
is close to $a_2(C)$ we extracted from a seemingly very different process of inclusive electroproduction\footnote{It took us nearly two years  to perform the analysis since we did not manage 
to convince  experimentalists to take the  deuteron and $^3$He target  data for exactly the same kinematics as for                        heavier nuclei and not to rely on the data taken in slightly different kinematics. This correction was very important due to the very fast variation of the cross section with $x$ and $Q^2$.}.

Another useful quantity, which  enters,  for example,  in the description of the high-energy backward nucleon production,  is a decay function~\cite{FS88} --- the probability to find  a nucleon with momentum  $p_1$ in the final state, if a nucleon with momentum $p_2$ is removed in a large energy-momentum process.
It should be rather close to the spectral function of the nucleus, when scattering off SRC takes place, although  in the case of the spectral function,  the motion of the pair may somewhat differ due to a difference in the interaction of the pair with the $A-2$  system.  
The resulting  picture of interaction in this model  is the scattering of a $\gamma^{\ast}(W)$  probe off a nucleon with momentum $\vec{p_1}$  destroying the binding  between  nucleons ``1'' and ``2'' and leading to the knockout of nucleons with momentum $p_1+ \vec q$ and production of a second nucleon with momentum  $\vec{p_2}\sim -\vec{p_1} $. The total momentum of the pair in the initial state  is small and can be estimated in the mean-field approximation~\cite{FS88,Frankfurt:1993sp}.

The simplest correlation effect in neutrino--nucleus scattering  is the shift of the $x$ distribution, when a backward proton was detected ---  detection of nucleons $\alpha$ and shift of the $x$ distributions. This is a kind of a Doppler effect as neutrino preferentially interacts with nucleons moving along the beam
\begin{equation}
\left<x_\alpha \right> =  (2 -\alpha) \left<x \right> \,.
\label{nuanti}
\end{equation}
Two experimental studies were performed using $\nu (\bar{\nu})$ beams. The second one had much higher statistics and used both deuteron and neon  fills~\cite{nu}.

\newpage
The data for the deuteron target are consistent with Eq.~(\ref{nuanti}), while for the neon target correlations are a factor of two weaker, which may be due to a contribution of secondary interactions.

An important experiment, EVA, was performed at BNL. Two protons were  detected in the forward direction in  at  back-to-back $p_\mathrm{t} $ corresponding to scattering near $90^\circ_\mathrm{cm}$. The aim was to look for the color transparency effect. There was enough  open space in the backward direction and  at some point, the Tel Aviv University group joined the experiments and  installed their  neutron detectors in the backward area.
Analysis of~\cite{Farrar} has demonstrated that due to a fast decrease of elastic $pp$ cross section with $s$ at fixed $\theta_\mathrm{cm}$, one selects events where the hit proton has a significant forward momentum leading to the production of a spectator backward  (this kinematics is similar to that of $(e,e'pn)$ reaction at $x<1$). The data analysis has shown that in $\sim 30 \%$  of events, a backward nucleon with a momentum $\>250$ MeV/$c$ is observed. Next, Misak Sargsian developed  Monte Carlo in which a decay function was modeled similar to the case of spectral function discussed above, light-cone dynamics was implemented  as well as final-state interactions. The analysis was published in 2006~\cite{Eli}. It was concluded that neutrons are emitted practically in all events with forward $pp$ trigger, though the motion of the pair and acceptance of the detector reduce the coincidence rate. Hence, it was concluded that the probability of $pp$  SRC is much smaller than of $np$ SRCs. This is consistent with the expectation of dominance of tensor SRCs.

These studies  were followed up with a spectacular series of experiments in  which the decay of  a nucleus  after the removal of a fast nucleon was studied. Selecting  different kinematics --- knock out of forward
(backward) going proton or neutron  allowed to test  self-consistency of the description of the process --- removal of a proton (neutron),  which in the initial state was moving along/opposite to $\vec{q}$.  For a detailed review, see~\cite{RMP}.
It is highly desirable to test the implementation of the f.s.i.\ which enters at  the energies too low for application of the Glauber model, \textit{etc}.

There are a number of fundamental issues that would have to be addressed including:
\begin{itemize}\parskip=-.5pt
\item[---] What is an effective  way  for probing $N \geq 3$ SRC? So far, the only way to observe them was inclusive backward proton production  at $\alpha_p \geq2$ --- in a  few nucleon SRC model~\cite{FS81} . The  highest $ \alpha$   for which 
protons were observed so far  is $\alpha_p\sim 3$.  
 To  describe the current data on the $ p A \to p +X$  reaction at the highest 
$\alpha $, one needs $N$ up to 4~\cite{FS81}\footnote{In the case of collision of two nuclei with $E_A/A  = 100$~GeV, $\alpha= 3$ corresponds to emission of protons with $E_p=300$~GeV.}.
 Indications of a set in of the $3N$ dominance were reported for the $A(e,e^\prime)$  reaction for  $x \geq 1.6$~\cite{Day23}.
\item[---] 
With the deuteron being an analog of the Born atom for quantum mechanics,  a broad range of high-momentum transfer reactions should be studied using different projectiles.
 \item[---]
Tests of factorization should be performed using lepton-- and hadron--deuteron scattering.   
\end{itemize}

In conclusion, a  significant progress has been  %was reached 
made in the study of short range correlations in the last decade.
Further progress in the field would be possible with the incorporation of constraints due to chiral dynamics on the scale of ${\mit\Delta}$ and nucleon mass splitting,   one of  the favorable areas of research of Mitya, Vitya, and Maxim, and additional constraints on
the nuclear dynamics from the experiments using electron and hadron beams at beam energies 10--30 GeV,

\vspace{7mm}
The research of M.S.\ was supported by the U.S.\ Department of Energy, Office of Science, Office of Nuclear Physics under Award No.~DE-FG02-93ER40771.

\flushleft


\begin{thebibliography}{99}


\bibitem{Bethe:1971xm}
H.A.~Bethe, {\em Theory of Nuclear Matter},  Annu. Rev. Nucl. Part. Sci.~21~93(1971)10.1146/annurev.ns.21.120171.000521~.

\bibitem{Levin}
E.M.~Levin, L.L.~Frankfurt, {\em The quark hypothesis and relations between cross-sections at high-energies},  JETP Lett.~2~65(1965)~.

\bibitem{Migdal}
A.B.~Migdal, {\em The Momentum Distribution of Interacting
Fermi Particles},  JETP~32~399(1957);
{Theory of Finite Fermi Systems and Applications to Atomic Nuclei} Wiley Interscience, New York 1967,10.1119/1.1975177~. 

\bibitem{DY}
D.M. Alde \etal, {\em Nuclear dependence of dimuon production at 800 GeV},  Phys. Rev. Lett.~64~2479(1990)10.1103/PhysRevLett.64.2479~.

\bibitem{West}
G.B. West, {\em Total neutron cross sections may not be what they seem to be},  Phys. Lett. B~37~509(1971)10.1016/0370-2693(71)90358-3~.

\bibitem{Westcom}
L.L.~Frankfurt, M.I.~Strikman, {\em Comment on the west correction to the total cross section},  Phys. Lett. B~64~433(1976)10.1016/0370-2693(76)90114-3~.

\bibitem{FS81}
L.L.~Frankfurt, M.I.~Strikman, {\em High-energy phenomena, short range nuclear structure and QCD},  Phys. Rep.~76~215(1981)10.1016/0370-1573(81)90129-0~.

\bibitem{FS88}
L.L.~Frankfurt, M.I.~Strikman, {\em Hard nuclear processes and microscopic nuclear structure},  Phys. Rep.~160~235(1988)10.1016/0370-1573(88)90179-2~.

\bibitem{core}
L.L.~Frankfurt, M.I.~Strikman, {\em New Direct Way of Checking the Nuclear Core Hypothesis in Inclusive Hadron Scattering Off the Polarized Deuteron},  Phys. Lett. B~76~285(1978)10.1016/0370-2693(78)90788-8~;
\textit{Erratum}  ibid.~80~433(1979)10.1016/0370-2693(79)91207-3~.

\bibitem{Weinberg}
S. Weinberg, {\em Dynamics at Infinite Momentum},  Phys. Rev.~150~1313(1966)10.1103/PhysRev.150.1313~.

\bibitem{modern}
L.L. Frankfurt, M.I. Strikman, {\em Short range correlations in nuclei as seen in hard nuclear reactions and light-cone dynamics}, in: eds.~{B. Frois, I. Sick}, 645-694,  {\em Modern Topics In Electron Scattering}, World Scientific~~1991.

\bibitem{Lech}
L.L. Frankfurt, M.I. Strikman, L. Mankiewicz, M. Sawicki, {\em Angular-momentum constraints in the light-cone quantum mechanics of the nucleon--nucleon system},  Few-Body Syst.~8~37(1990)10.1007/BF01079801~.

\bibitem{Kim}
K.V.~Alanakian \etal, {\em Emission of cumulative protons in the reaction $^{12}$C($e,e'p$)},  Phys. Atom. Nucl.~61~207(1998)~ [ Yad. Fiz.~61~256(1998)~]. 

\bibitem{Stavinsky}
V.S. Stavinsky, {\em Limiting Fragmentation of Nuclei --- the Cumulative Effect},  Fiz. Elem. Chast. Atom. Yadra~10~949(1979)~, in Russian.

\bibitem{Leksin}
Yu.D.~Bayukov \etal, {\em Angular Dependences of Inclusive Nucleon Production in Nuclear Reactions at High-energies and Separation of Contributions from Quasifree and Deep Inelastic Nuclear Processes},  Sov. J. Nucl. Phys.~42~116(1985)~ and references therein.  

\bibitem{2Ndisc}
L.L. Frankfurt, M.I. Strikman, {\em Pion-exchange contributions to the isoscalar current and magnetic moment operators},  Phys. Lett. B~69~51(1977)10.1016/0370-2693(77)90130-7~.

\bibitem{CiofidegliAtti:1991mm}
C.~Ciofi degli Atti, S.~Simula, L.L.~Frankfurt, M.I.~Strikman, {\em Two-nucleon correlations and the structure of the nucleon spectral function at high values of momentum and removal energy},  Phys. Rev. C~44~R7(1991)10.1103/PhysRevC.44.R7~.

\bibitem{Frankfurt:1993sp}
L.L.~Frankfurt, M.I.~Strikman, D.B.~Day, M.~Sargsian, {\em Evidence for short range correlations from high $Q^2$ ($e, e^\prime$) reactions},  Phys. Rev. C~48~2451(1993)10.1103/PhysRevC.48.2451~.

\bibitem{nu}
E. Matsinos \etal , {\em Backward particle production in neutrino neon interactions},  Z. Phys. C~44~79(1989)10.1007/BF01548585~.

\bibitem{Farrar}
G.R. Farrar, H. Liu, L.L. Frankfurt, M.I. Strikman, {\em Transparency in Nuclear Quasiexclusive Processes with Large Momentum Transfer},  Phys. Rev. Lett.~61~686(1988)10.1103/PhysRevLett.61.686~.

% \bibitem{Eli}
% L.L.~Frankfurt, M.I.~Strikman, {\em New Direct Way of Checking the Nuclear Core Hypothesis in Inclusive Hadron Scattering Off the Polarized Deuteron},   Phys. Lett. B~76~285(1978)10.1016/0370-2693(79)91207-3~.

% \bibitem{Frankfurt:1978kw}
% L.L.~Frankfurt, M.I.~Strikman, {\em New direct way of checking the nuclear core hypothesis in inclusive hadron scattering off the polarized deuteron},  Phys. Lett. B~76~285(1978)10.1016/0370-2693(78)90788-8~;
% \textit{Erratum}  ibid.~80~433(1979)10.1016/0370-2693(79)91207-3~.

\bibitem{Eli}
E.~Piasetzky \etal , {\em Evidence for Strong Dominance of Proton--Neutron Correlations in Nuclei},  Phys. Rev. Lett.~97~162504(2006)10.1103/PhysRevLett.97.162504~.

\bibitem{RMP}
O. Hen, G.A. Miller, E. Piasetzky, L.B. Weinstein, {\em Nucleon--nucleon correlations, short-lived excitations, and the quarks within},  Rev. Mod. Phys.~89~045002(2017)10.1103/RevModPhys.89.045002~.

\bibitem{Day23}
M. Sargsian, D.B. Day, L.L. Frankfurt, M. Strikman, {\em Toward observation of three-nucleon short-range correlations in high-$q^2$ $A(e,e^\prime)X$ reactions},  Phys. Rev. C~107~014319(2023)10.1103/PhysRevC.107.014319~.


\end{thebibliography}
\end{document}